\documentclass[prl,amsmath,twocolumn,amssymb,aps,superscriptaddress,floatfix,preprintnumbers,10pt,notitlepage]{revtex4-1}
\RequirePackage[switch,columnwise]{lineno}
\usepackage{graphicx}
\usepackage{amsmath, braket, amsfonts}
\usepackage{amssymb}
\usepackage{natbib}
\usepackage{sidecap}
\usepackage{bm, color, ulem}
\usepackage{comment}
\usepackage{xcolor}
\usepackage{tikz}
\usepackage{pgfplots}
\usepackage{siunitx}
\pgfplotsset{compat=1.12}
\definecolor{ferngreen}{rgb}{0.31, 0.47, 0.26}

\begin{document} 
\title{Imaging the breakdown of ohmic transport in graphene}

\author{A. Jenkins}
\affiliation{Department of Physics, University of California, Santa Barbara CA 93106, USA}

\author{S. Baumann}
\affiliation{Department of Physics, University of California, Santa Barbara CA 93106, USA}

\author{H. Zhou}
\affiliation{Department of Physics, University of California, Santa Barbara CA 93106, USA}

\author{S. A. Meynell}
\affiliation{Department of Physics, University of California, Santa Barbara CA 93106, USA}

\author{D. Yang}
\affiliation{Department of Physics, University of California, Santa Barbara CA 93106, USA}

\author{K. Watanabe}
\affiliation{National Institute for Materials Science, 1-1 Namiki, Tsukuba 305-0044, Japan}

\author{T. Taniguchi}
\affiliation{National Institute for Materials Science, 1-1 Namiki, Tsukuba 305-0044, Japan}

\author{A. Lucas}
\affiliation{Department of Physics and Center for Theory of Quantum Matter, University of Colorado, Boulder CO 80309 USA}

\author{A. F. Young}
\email{andrea@physics.ucsb.edu}
\affiliation{Department of Physics, University of California, Santa Barbara CA 93106, USA}

\author{A. C. Bleszynski Jayich}
\email{ania@physics.ucsb.edu}
\affiliation{Department of Physics, University of California, Santa Barbara CA 93106, USA}

\date{\today}

\begin{abstract}

Ohm's law describes the proportionality of current density and electric field.  In solid-state conductors, Ohm's law emerges due to electron scattering processes that relax the electrical current.  Here, we use nitrogen-vacancy center magnetometry to directly image the local breakdown of Ohm's law in a narrow constriction fabricated in a high mobility graphene monolayer.  Ohmic flow is visible at room temperature as current concentration on the constriction edges, with flow profiles entirely determined by sample geometry.  However, as the temperature is lowered below 200 K, the current concentrates near the constriction center.  The change in the flow pattern is consistent with a crossover from diffusive to viscous electron transport dominated by electron-electron scattering processes that do not relax current.
\end{abstract}

\maketitle

\newpage

Ohm's law states that the current flow through an electrical conductor is proportional to a voltage difference across it.  While this introductory textbook physics is ubiquitous in macroscopic electrical devices, Ohm's law need not hold locally at every point inside of a conductor.  Specifically, Ohm's law arises only on length scales sufficiently long that microscopic scattering processes completely relax the electrical current.  In an ordinary metal, impurity scattering and umklapp processes, each of which relax current, dominate the electronic dynamics; hence Ohm's law arises on scales larger than the electronic mean free path, $\ell_{\mathrm{mr}}$. In this regime, electrical transport is diffusive. 

However, in low-density, low-disorder Fermi liquids, it was predicted decades ago that dynamics could be dominated by electron-electron collisions, which conserve momentum.  In the regime where the electron-electron scattering length $\ell_{\mathrm{ee}}\ll \ell_{\mathrm{mr}}$, the momentum-conserving collisions do not completely relax the electrical current, resulting in viscous rather than diffusive transport, with current flow resembling that of a fluid \cite{gurzhi_hydrodynamic_1968}. 
Following preliminary work in the 1990s on III-V semiconductor heterostructures \cite{DeJong1995a}, a slew of electrical and thermal transport anomalies observed in clean graphene heterostructures have been linked to the onset of 
electron hydrodynamics \cite{Crossno2016,KrishnaKumar2017,gallagher_quantum-critical_2019,berdyugin_measuring_2019,bandurin_fluidity_2018,bandurin_negative_2016}.  Most recently, imaging studies of the Hall voltage in a small magnetic field have revealed a crossover with rising temperature from a ballistic regime, where voltage is out of equilibrium, to a regime of strong voltage equilibration \cite{Sulpizio2019a}, interpreted as evidence for a viscous regime.  However, the breakdown of ohmic transport due to the onset of electron-electron dominated scattering has not been definitively observed.  

Here we use direct imaging of the current flow profile to observe the local breakdown of Ohm's law in a monolayer graphene device in which a narrow constriction has been etched. Using a scanning nitrogen vacancy (NV) center magnetometer, we image the local magnetic field above the device, related by the Biot-Savart law to the current flow profile through the constriction. These current profiles are expected to be different if the flow is limited by impurities (ohmic), electron-electron collisions (hydrodynamic), or boundary scattering (ballistic) \cite{Guo2017, KrishnaKumar2017, Guo2018a}. In the ohmic regime, current concentrates near the constriction boundaries, mathematically equivalent to the bunching of the electric field lines near the corners of a lightning rod. Hence, the presence or absence of the current bunching provides a clear means of distinguishing the transition.  In our experiment, the relevant ratios between $\ell_{\mathrm{ee}}$,  $\ell_{\mathrm{mr}}$, and the constriction width $w$ can be tuned \textit{in situ} via control of the carrier density and temperature.     
Our measurements definitively resolve the dramatic transition from ohmic to non-ohmic flow and provide evidence of a robust collision-dominated regime in which $\ell_{\mathrm{ee}} < \ell_{\mathrm{mr}}$. 

\begin{figure*}
\begin{centering}
\includegraphics{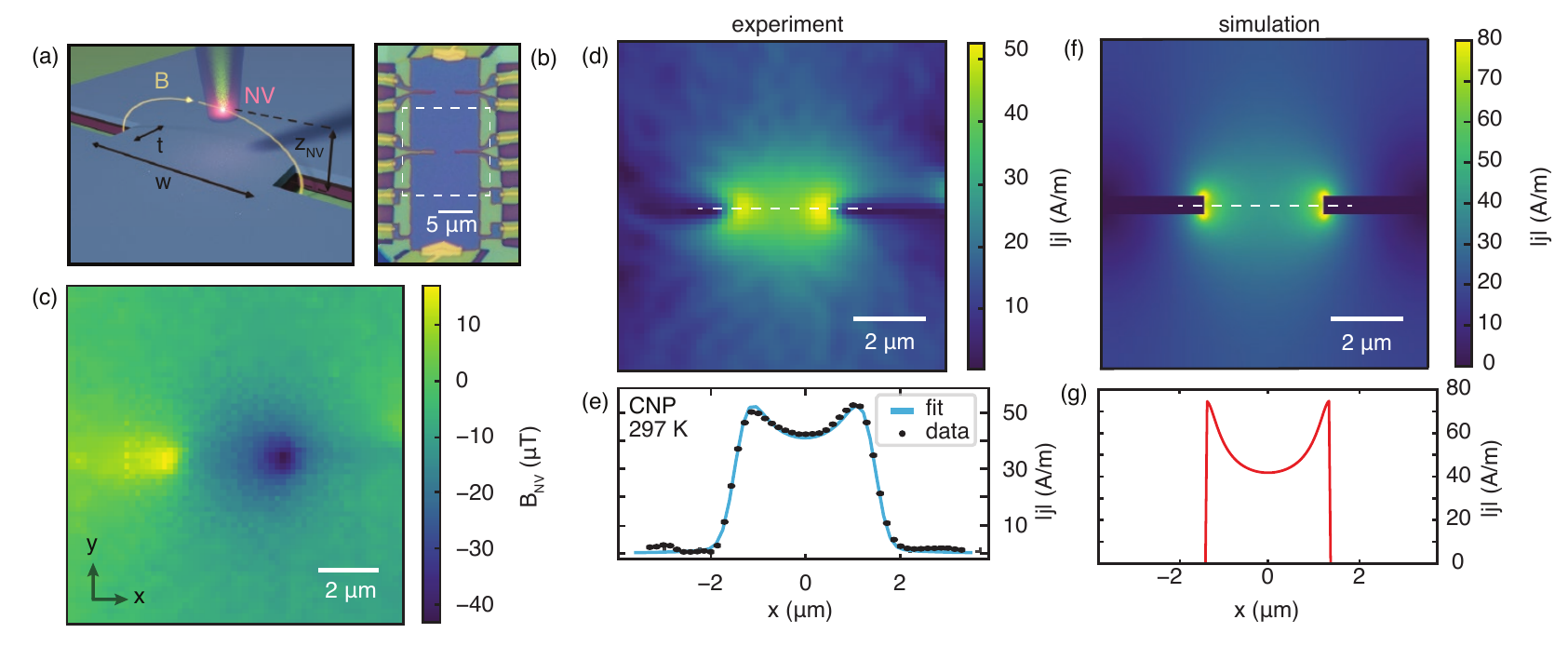}
\caption{\textbf{Room temperature current imaging at the charge neutrality point.} 
\label{fig:figExpIntro} (a) Experimental setup. A diamond probe containing a $\SI{30}{\nano\meter}$-deep, single NV center is scanned over hBN-encapsulated monolayer graphene device while current flows through an etched constriction. The stray magnetic field produced by the flowing current is measured via shifts in the NV magnetic resonance spectrum and the measured field is used to reconstruct the underlying current distribution. Optical image of the graphene device. (c) Scanning NV magnetometry signal over the area indicated by the white dashed line in (b) at room-temperature and at the charge neutrality point. The image boundaries, where the field varies slowly, are measured with sparser sampling and linear interpolation is used to fill in these regions (see supplementary information). (d) Reconstructed current density magnitude $|j|$ at 298 K and carrier density near the CNP ($n<\SI{0.06e12}{\centi\meter^{-2}}$) (e) A linecut of $|j|$ taken along the dashed white line in d). The solid line plots the expected current profile, obtained using a parameter-free, purely ohmic model and shows good agreement with the data. This model takes into account broadening of the features in $|j|$ due to the finite distance between the NV and the graphene. (f) The simulated ohmic current in the device without the NV filter function applied. (g) A linecut taken along the the dashed white line in f) displays a pronounced double-peak feature indicative of ohmic transport
.}
\end{centering}
\end{figure*}

\begin{figure*}[t]
\begin{centering}
\includegraphics{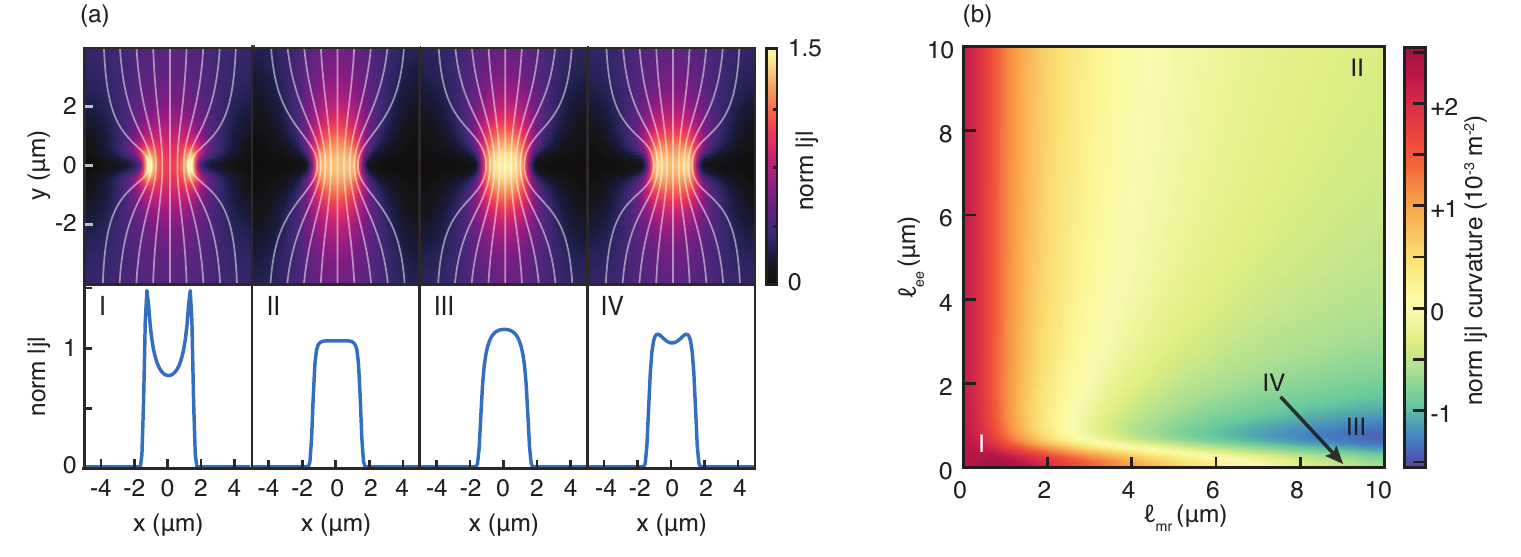}
\caption{\textbf{Boltzmann transport model.}
\label{fig:figTheoryIntro} (a) Four examples of current flow profiles for a slit of finite thickness and a width, $w = \SI{3}{\micro\meter}$ and thickness $t = \SI{0.4}{\micro\meter}$. The free parameters in the model are $\ell_{\mathrm{ee}}$ and $\ell_{\mathrm{mr}}$. The current density is normalized by the average current density through the slit. Line profiles of the current across the center of the slits are shown underneath. Region I corresponds to the ohmic regime where $\ell_{\mathrm{mr}} = \ell_{\mathrm{ee}} \approx 0$. Region II is in the deep ballistic regime with $\ell_{\mathrm{mr}} = \ell_{\mathrm{ee}} = \SI{10}{\micro\meter}$. The hydrodynamic regime is represented by III ($\ell_{\mathrm{ee}} = \SI{1}{\micro\meter}$, $\ell_{\mathrm{mr}} = \SI{10}{\micro\meter}$)  where the current shape assumes a more elliptical profile. A final region, IV ($\ell_{\mathrm{ee}} \approx 0$, $\ell_{\mathrm{mr}} = \SI{10}{\micro\meter}$, corresponds to when electron-electron scattering becomes so dominant that it begins to reduce the effective current relaxation length and results in viscous bunching near the slit edges. (b) The curvature of the current, $\mathrm{d}^2 |j|/\mathrm{d}x^2$, at the center of the slit, normalized by the average current density through the slit. The changes in profiles in the regimes of II and III are subtle to resolve when uncertainty in the slit width is taken into account.}
\end{centering}
\end{figure*}

Figure \ref{fig:figExpIntro}a-b show a schematic diagram of our experimental setup \cite{pelliccione_scanned_2016} and optical image of the graphene device.  Current ($I = \SI{150}{\micro A}$) is passed through a graphene constriction of nominal width $w = \SI{3}{\micro\meter}$ and thickness $t = \SI{0.4}{\micro\meter}$ etched in a monolayer of high-quality graphene encapsulated by hexagonal boron nitride. The flowing current produces a magnetic field that is sensed by the NV center. The magnetic resonance signal of the NV center is detected optically as the sample is scanned with respect to the NV at a constant height, with NV-graphene separations ranging from $z_{\mathrm{NV}}$=140--170 nm between different scans. 
The NV axis is oriented with a polar angle of $\theta=55^\circ$ relative to the device normal and an azimuthal angle of $\phi=173^\circ$ relative to the image $x$-axis. 
The resulting spatial map of the stray magnetic field (Figure \ref{fig:figExpIntro}c) is then converted using standard Fourier domain techniques \cite{Roth1989, Chang2017} into the total current density ($|j|$) in the vicinity of the constriction (see Figure \ref{fig:figExpIntro}d-e and Ref. \cite{Note1}).

We first image room temperature current flow at the charge neutrality point, shown in Figs. \ref{fig:figExpIntro}d-e. Here graphene behaves as a non-Fermi liquid, with electrical transport dominated by current-relaxing recombination of thermally excited electrons and holes. The current profile is thus expected to be ohmic \cite{lucas_hydrodynamics_2018}. 
Fig. \ref{fig:figExpIntro}e shows the total reconstructed current density $|j|$ across the graphene constriction. The current profile shows distinctive peaks near the constriction boundaries, consistent with expectations for ohmic transport. Indeed, the data are quantitatively well matched by a parameter-free fit to an ohmic model that assumes a spatially uniform local conductivity, whose value is independently measured with transport. 
It bears noting the the finite distance between NV center puts a fundamental limit on spatial resolution.  The reconstructed $|j|$ in Fig. \ref{fig:figExpIntro}d-e is thus related to the physical current density via a spatial low pass filter. Figure \ref{fig:figExpIntro}f-g show the expected current distribution used to generate the fit in Fig. \ref{fig:figExpIntro}e. 
We conclude that at room temperature and at the CNP, transport through our device is diffusive, and momentum conserving electron-electron scattering is unimportant.

In order to understand the qualitative behavior of current flow away from the ohmic regime, we perform simulations of the quantum Boltzmann equation (QBE)\cite{Note1} in our constriction geometry, which allow us to capture the effects of finite $\ell_{\mathrm{mr}}$ and $\ell_{\mathrm{ee}}$.
Figure \ref{fig:figTheoryIntro}a shows simulation results assuming $w = \SI{3}{\micro\meter}$ for varying $\ell_{\mathrm{ee}}$ and $\ell_{\mathrm{mr}}$. 
Evidently, current flow profiles can vary dramatically between transport regimes; in particular, the strong current concentration on the slit boundary is specific to the ohmic regime. To capture these qualitative differences in a single quantitative figure of merit, Fig. \ref{fig:figTheoryIntro}b shows the curvature of $|j|$ at the center of the slit plotted
as a function of $\ell_{\mathrm{ee}}$ and $\ell_{\mathrm{mr}}$. 
Deep in the ohmic regime (I) where $\ell_{\mathrm{mr}} \ll w,\ell_{\mathrm{ee}}$, the expected flow profile shows good agreement with the data and ohmic model fit of Fig. \ref{fig:figExpIntro}g. 
The ballistic regime (II) is the limit of $w \ll \ell_{\mathrm{mr}},\ell_{\mathrm{ee}}$, and results in a flat current profile. Hydrodynamic effects become visible when momentum conserving collisions are dominant and the momentum relaxing length scale is large compared to the device size, $\ell_{\mathrm{ee}}\ll w < \ell_{\mathrm{mr}}$  (III). 
At extremely small $\ell_{\mathrm{ee}} \ll w^2/\ell_{\mathrm{mr}}$ there exists a final regime (IV), deep within the hydrodynamic limit, where viscous effects are so strong that a weakly double-peaked current profile again arises, as interactions reduce the effective momentum relaxation length to $\sqrt{\ell_{\mathrm{ee}}\ell_{\mathrm{mr}}}$ \cite{lucas_hydrodynamics_2018}.
In a realistic device and measurement geometry, the ballistic and weakly hydrodynamic regions (II and III) exhibit current maxima near the constriction center and can be difficult to distinguish: the subtle quantitative differences in boundary profiles are easily obscured by the finite spatial resolution arising from the finite distance between current and magnetic field sensor.

\begin{figure*}[t]
\begin{centering}
\includegraphics{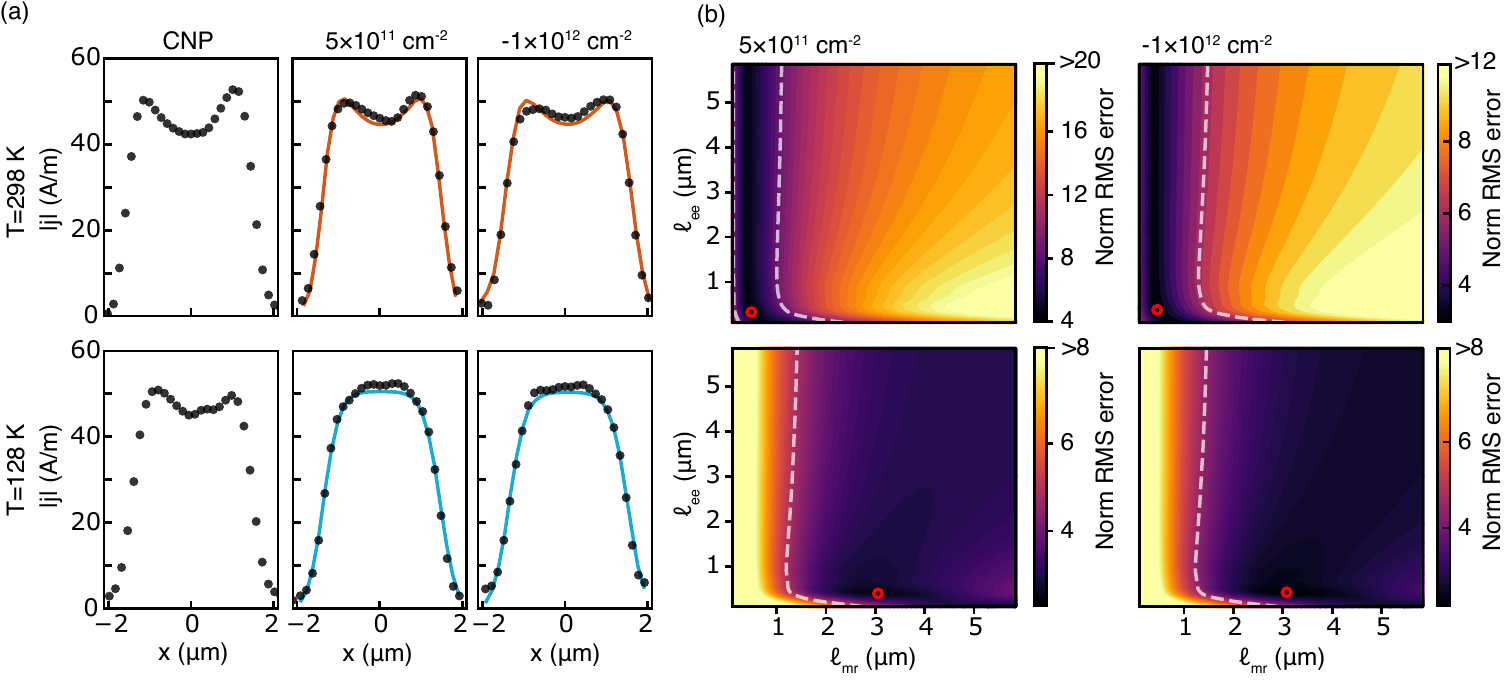}
\caption{\textbf{Local breakdown of ohmic transport.}
\label{fig:fig3}  (a) Flow profiles across the center of the slit at room temperature (top row, red) and at $T = \SI{128}{\kelvin}$ (bottom row, blue). At room temperature, ohmic double-peaked current profiles are always visible, indicating the importance of momentum-relaxing scattering processes.  $\ell_{\mathrm{mr}}$ can be tightly constrained at room temperature, and the current flow is insensitive to $\ell_{\mathrm{ee}}$ in the regime $\ell_{\mathrm{ee}}>500$ nm. For lower temperatures, at finite density we observe the breakdown of the double-peaked profile and observe a rounder, single-peaked profile -- the qualitative signature of the absence of strong momentum-relaxing scattering. (b) Maps of the RMS error per pixel between QBE simulations and the reconstructed current densities, normalized by the current reconstruction noise. At room temperature, the fits are consistent with diffusive transport and inconsistent with a hydrodynamic or ballistic flow. For lower temperatures, our data is inconsistent with both the deeply hydrodynamic regime and inconsistent with the diffusive regime - an indication that the device sits near the boundary between the weakly hydrodynamic and ballistic regimes. The white dashed lines correspond to fits where the RMS error is twice that of the best fit. We take the areas bounded from above by these contours as consistent with the experimental data. The red dot corresponds to the absolute best fit in these data.}
\end{centering}
\end{figure*}

Presented in Fig. \ref{fig:fig3}a are the experimentally measured current flow profiles at several carrier densities, including charge neutrality, at both $T = \SI{298}{\kelvin}$ and $T = \SI{128}{\kelvin}$.  
At room temperature, the double-peaked current profile persists as the density is tuned away from charge neutrality.  However, the size of the peaks is reduced with density, consistent with $\ell_{\mathrm{mr}}$ \textit{increasing} with density as expected from transport theories of monolayer graphene \cite{DasSarma}. 
At $T = \SI{128}{\kelvin}$ and at the CNP, we again observe the double-peaked profile, indicating that charge recombination dominates scattering near the CNP even at low temperatures. However, at finite densities and $T = \SI{128}{\kelvin}$, we observe the emergence of a single-peaked current profile, an unambiguous demonstration of locally non-ohmic transport. 

Because both ballistic and hydrodynamic transport are expected to be peaked in the center of the constriction, we turn to a quantitative comparison of our experimental data to the QBE simulations described above to place bounds on the possible values of $\ell_{\mathrm{ee}}$ and $\ell_{\mathrm{mr}}$ in our device.  
Simulations and measurements are compared over a two dimensional area of dimensions $4.0 \times 1.6$ microns centered on the constriction (see \cite{Note1}). 
Figure \ref{fig:fig3}b shows the root mean square (RMS) residual per pixel of the fit normalized by the standard deviation of the current reconstruction noise, plotted as a function of the simulation input parameters $\ell_{\mathrm{ee}}$ and $\ell_{\mathrm{mr}}$.  The white dashed lines are contours corresponding to twice the RMS residuals of the best fit. We take areas of the parameter space bounded from above by these contours (darker shades in Fig. \ref{fig:fig3}b) as consistent with experimental data.

At room temperature, magnetometry data constrain $\ell_{\mathrm{mr}} \lesssim \SI{1}{\micro m}$, while providing no constraint on the value of $\ell_{\mathrm{ee}}$.  This is consistent with estimates for scattering by momentum-relaxing phonons, which dominate scattering on experimentally relevant length scales resulting in ohmic flow. 
Best fit regions are markedly different at $T = \SI{128}{\kelvin}$, and most consistent with the crossover regime intermediate between hydrodynamic and ballistic flow. At T = $\SI{128}{\kelvin}$, the average RMS residual is minimal for $\ell_{\mathrm{ee}} = \SI{0.4}{\micro m}$ and $\ell_{\mathrm{mr}} = \SI{3.1}{\micro m}$ for both $n=\SI{5e11}{/\centi\meter^2}$ and $n=\SI{-1e12}{/\centi\meter^2}$. However, larger values of $\ell_{\mathrm{ee}}$ and $\ell_{\mathrm{mr}}$ are only weakly constrained, primarily due to uncertainty in the precise nature of the boundary conditions, as well as uncertainty in the constriction width.  An independent check on these fits can be obtained from measurements of the conductivity $\sigma$ in the same device (see \cite{Note1}), from which we extract the mean free path $\ell_{\mathrm{mfp}}=h\sigma/2k_{\mathrm{F}}e^2$, where $k_{\mathrm{F}}=\sqrt{\pi n}$ is the Fermi wavevector. 
In the Fermi liquid regime, $\ell_{\mathrm{mfp}}$ approximates $\ell_{\mathrm{mr}}$.  We find that at both $\SI{298}{\kelvin}$ ($\ell_{\mathrm{mfp}}\approx \SI{0.9}{\micro\meter}$) and $\SI{128}{\kelvin}$ ($\ell_{\mathrm{mfp}}\approx \SI{4.3}{\micro\meter}$) the measured mean free path agrees with the momentum relaxation length obtained from the NV magnetometry fits. Our data thus completely rule out both the strongly interacting hydrodynamic regime (IV) and the ohmic regime (I).

In order to conclusively resolve the subtle differences in flow patterns in the hydrodynamic/ballistic crossover regime, a geometry better suited to measuring these differences or a sharper understanding of the realized boundary conditions
will be required. 
However, our results, when taken together with those of Ref. \cite{Sulpizio2019a}, provide strong evidence that monolayer graphene hosts a weakly interacting hydrodynamic regime at intermediate temperatures, manifesting in strongly modified local current and voltage distribution patterns.
Looking forward, locally resolved current measurements may be useful in conclusively resolving hydrodynamic effects (or the lack thereof) in more exotic materials \cite{Moll2016, Gooth2018}.  More broadly, these techniques provide a powerful method for visualizing electronic dynamics normally invisible in bulk resistivity measurements.

\phantom{\footnote{Materials and methods are available as supplementary materials at the Science website.}}

\section*{Acknowledgments}
The authors acknowledge discussions with L. Levitov, and thank J. Sanchez-Yamagishi for comments on the manuscript.  
This work was primarily supported by the National Science Foundation under award DMR-1810544. 
KW and TT acknowledge support from the Elemental Strategy Initiative conducted by the MEXT, Japan and the CREST (JPMJCR15F3), JST.
AFY acknowledges the support of the David and Lucile Packard Foundation and the Alfred. P. Sloan Foundation. SAM acknowledges the support of the Natural Sciences and Engineering Research Council of Canada (NSERC), [AID 516704-2018].
\bibliography{graphene_imaging_f}

\bibliographystyle{Science.bst}

\clearpage
\onecolumngrid
\renewcommand{\figurename}{Fig. S}
\renewcommand{\theequation}{S\arabic{equation}}
\setcounter{equation}{0}
\setcounter{figure}{0}
\section*{Materials and Methods}
\subsection*{Device fabrication}
\begin{figure*}[h]
\begin{centering}
\includegraphics{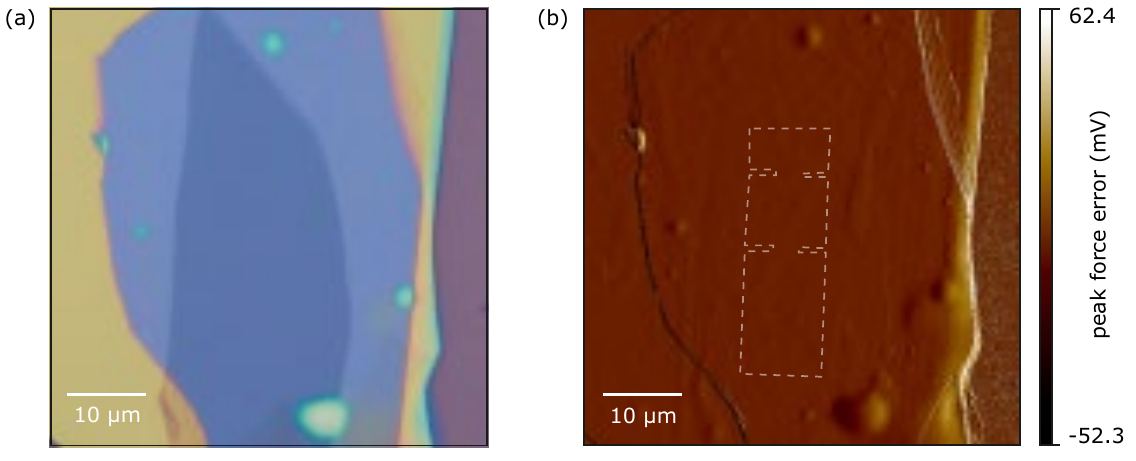}
\let\nobreakspace\relax
\caption{\label{fig:AFM} (a) Optical image of the graphene stack prior to fabrication of the device shown in the main text. (b) Peak force AFM error image used to locate bubbbles in the stack. The device is positioned in the bubble-free region at the position of the white dashed line.}
\end{centering}
\end{figure*}

The  device used in the main text of this work consists of a hexagonal-boron nitride (hBN) ecapsulated monolayer (ML) of graphene with a graphite gate used to tune the carrier density. The thicknesses of the upper and lower hBN layers are 99 nm and 50 nm respectively. This geometry was chosen to reduce scattering due to charge disorder\cite{wang_one-dimensional_2013,zibrov_tunable_2017}. The ML graphene, graphite gate and hBN flakes were prepared by mechanical exfoliation and the final heterostructure was fabricated using a dry transfer procedure \cite{Zomer2014}. Figure S\ref{fig:AFM} a-b show an optical and atomic force microscope image of the graphene heterostructure before we etched the device geometry. The device boundary was defined with plasma etching and a layer of Cr/Pd/Au was deposited to form the electrodes.

\subsection*{Magnetometry methods}
\begin{figure*}[h]
\begin{centering}
\includegraphics{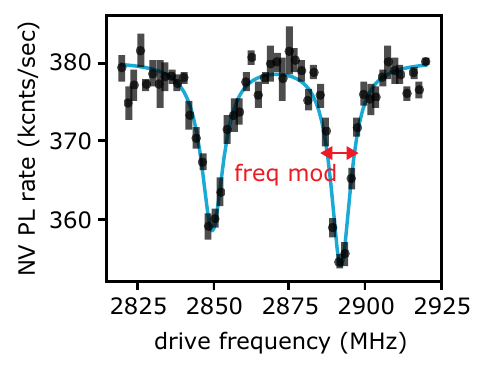}
\let\nobreakspace\relax
\caption{\label{fig:CWESR} Example measurement of the NV ESR spectrum. Dips in the ESR spectrum correspond to resonances of the applied microwaves with the $m_s =0 \rightarrow m_s=\pm 1$ spin transitions. The red arrow indicates the approximate size and location of the frequency modulation used during scanning measurements.}
\end{centering}
\end{figure*}

Figure S\ref{fig:CWESR} shows an example the ground state ESR spectrum of an NV. For measurements of the small stray fields produced by currents flowing in the graphene device, we apply a square-wave modulation at 5 kHz close to the estimated peak minimum and sweep the center frequency over a range of 0.6--1 MHz. The size of the amplitude of modulation is chosen relative to the width of the ESR spin transitions and was typically 6 MHz for the measurements described in this work. This technique is similar to that described in \cite{Schoenfeld}. The resulting demodulated PL signal has an x-intercept at the frequency of the NV spin transition. In the scanning measurements, we measure the NV frequency for both current directions $\pm I$ and the center frequency of the sweep range is updated for each direction at each point in the scan.

\begin{figure*}
\begin{centering}
\includegraphics{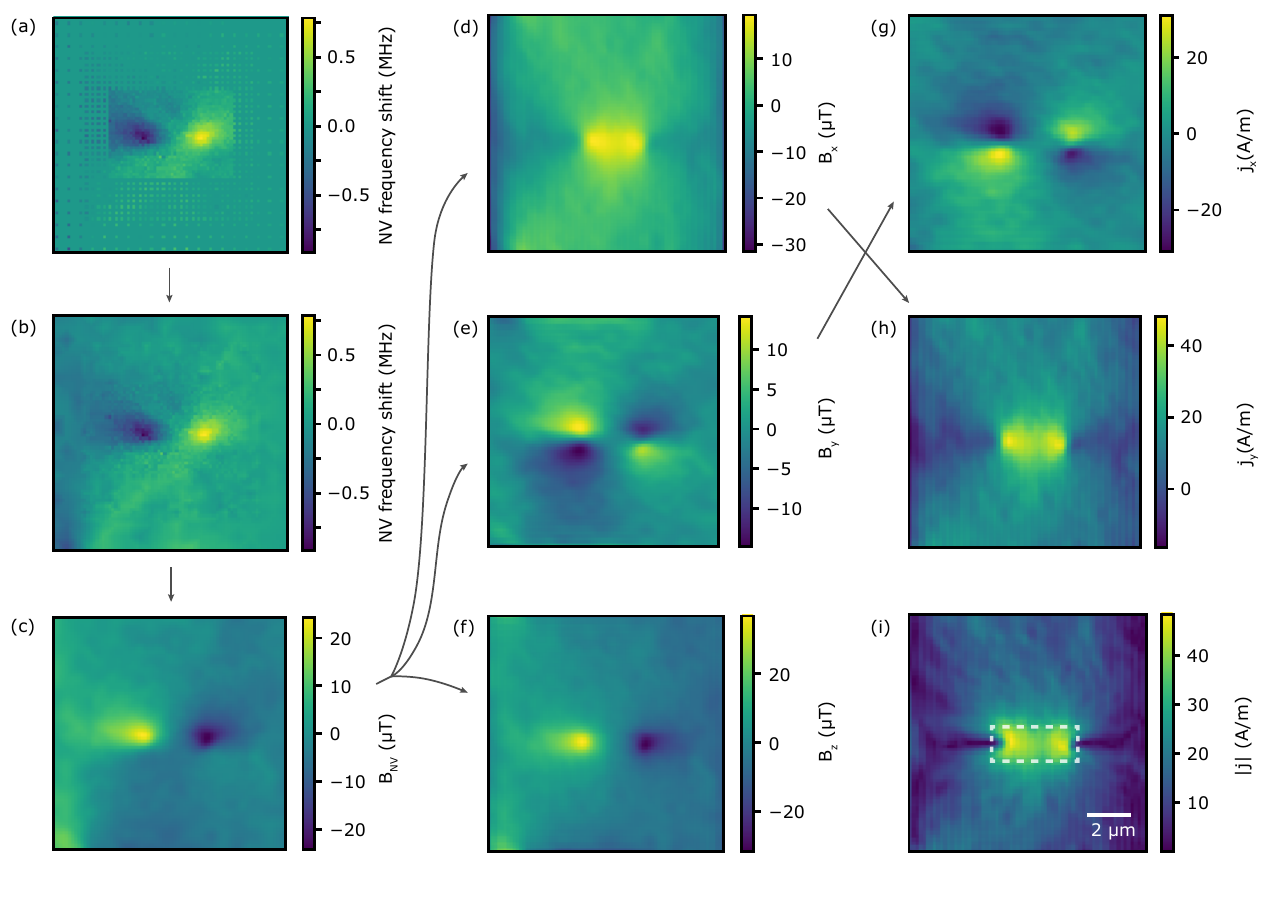}
\let\nobreakspace\relax
\caption{\label{fig:field_current_recon} Example of current reconstruction at 298 K, near the CNP. (a) The raw measured frequency of the NV spin transition, showing the area of fine field sampling with sparser spacing away from the constriction. (b) The frequency data with linear interpolation in the sparsely sampled areas. (c) The field along the NV axis with spatial averaging with a Gaussian filter with size equal to the NV-graphene separation (approximately the imaging resolution). (d-f) The vector components of the magnetic field calculated from (c). (g-i) Current density components $j_x$, $j_y$, and current density magnitude $|j|$ reconstructed from $B_x$ and $B_y$. The dashed line in (i) outlines the QBE fitting area of the image. The scale bar in (i) applies to all images.}
\end{centering}
\end{figure*}

\begin{figure*}
\begin{centering}
\includegraphics[width = 6 in]{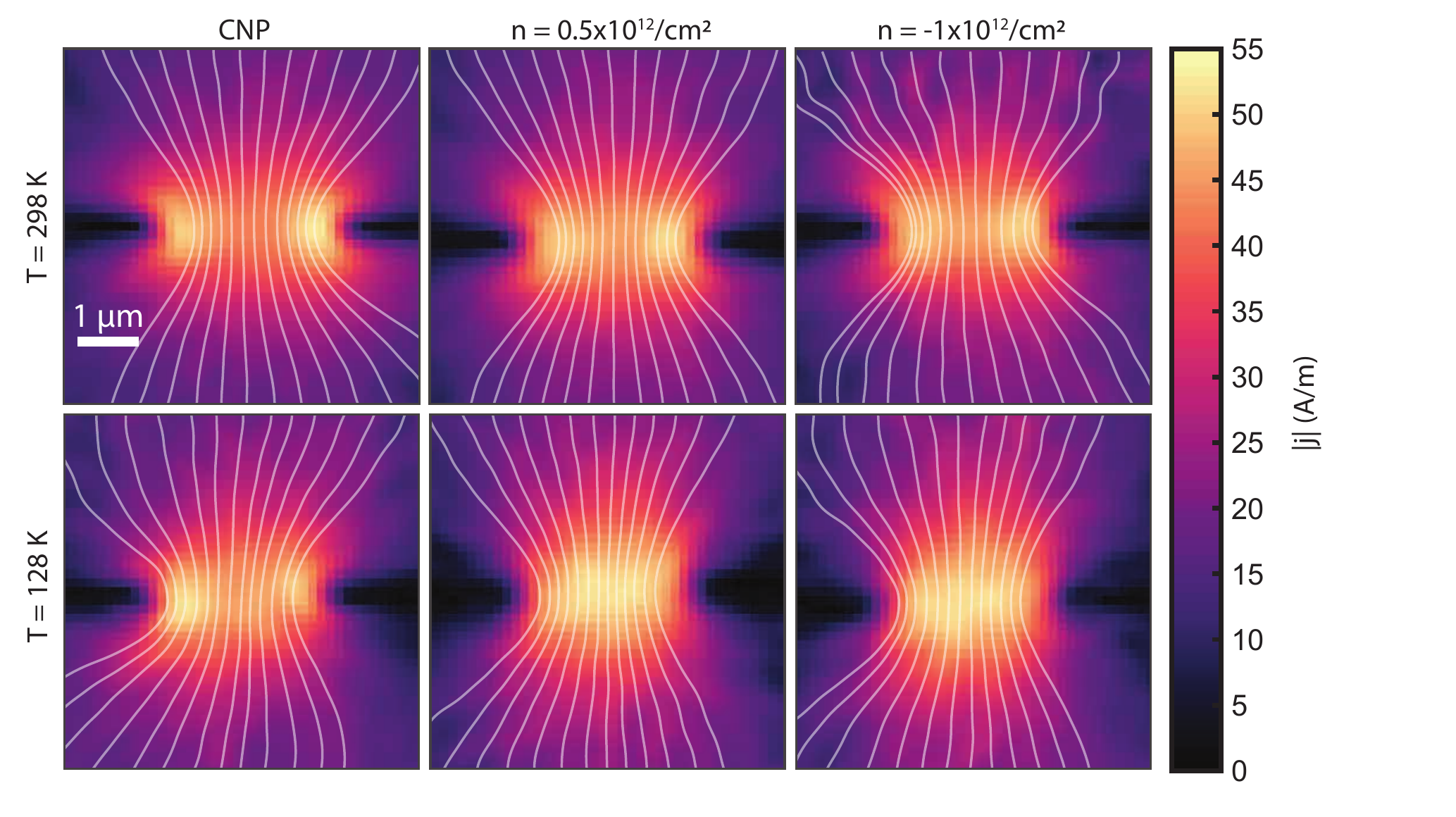}
\let\nobreakspace\relax
\caption{\label{fig:datastreamlines} Reconstructed currents for T = 297 K and T = 128 K for the carrier densities studied in this paper. The measured current has been reconstructed from field measurements using the process illustrated in Fig. \ref{fig:field_current_recon}.}
\end{centering}
\end{figure*}

We use adaptive sampling to measure the stray field with a higher resolution closer to the constriction relative to the boundaries of the image. An example of this adaptive sampling is shown in Fig. S\ref{fig:field_current_recon}a where the sampling is chosen to be fine relative to our resolution where the field is expected to be quickly varying and is chosen to be sparse everywhere else. We fill in the sparse areas of the image with linear interpolation to reconstruct the current density, as shown in Fig.S\ref{fig:field_current_recon}b. We average the over-sampled frequency map with a Gaussian filter with a width determined by the NV-graphene separation to get the field along the NV axis as in Fig. S\ref{fig:field_current_recon}c. In the magnetostatic limit, this single component of magnetic field can be used to compute the full vector magnetic field as in \cite{Lima2009}. In Fourier space, the components are
\begin{align}\label{eqn:Bnv_recon}
&\tilde{B}_{\mathrm{NV}} =  \left(-\frac{\mathrm{i} k_x}{k} \sin(\theta)\cos(\phi) - \frac{\mathrm{i} k_y}{k} \sin(\theta) \sin(\phi) + \cos(\theta) \right) \tilde{B}_z
\end{align}
\begin{align}\label{eqn:Bx_recon}
&\tilde{B}_x(k_x, k_y, z)  = -\frac{\mathrm{i} k_x}{k}\tilde{B}_z(k_x, k_y, z)
\end{align}
\begin{align}\label{eqn:By_recon}
&\tilde{B}_y(k_x, k_y, z)  = -\frac{\mathrm{i} k_y}{k}\tilde{B}_z(k_x, k_y, z)
\end{align}
For two dimensional sheets of current, these components of stray field can be mapped directly to current densities for spatial frequency components $k<1/z_{\mathrm{NV}}$. In Fourier space the fields and current densities are related by \cite{Roth1989, Chang2017},
\begin{align}
&\mathcal{F}[j_y] = \frac{2 \mathrm{e}^{k z}}{\mu_0} \mathcal{F}[B_x]\\
&\mathcal{F}[j_x] = - \frac{2 \mathrm{e}^{k z}}{\mu_0}  \mathcal{F}[B_y]
\end{align}
for NV height $z$ and wavenumber $k$. As described previously, we limit the current reconstruction to $k<1/z_{\mathrm{NV}}$ (using a Gaussian filter with width $\sigma = \sqrt{2} z_{\mathrm{NV}}$) to avoid the exponential amplification of high frequency noise. Another limitation of the reconstruction is that the $k=0$ components of both $j_x$ and $j_y$ can not be extracted from the field measurement alone. To set the zero level of current density, we shift each reconstructed component so that the current density inside the etched area of the constriction is zero. We then fit the magnitude of the reconstructed current density to the QBE using the area outlined in Fig. S\ref{fig:field_current_recon}i by a dashed white line. The reconstructed data for the device described in the paper are shown in Fig. \ref{fig:datastreamlines}.

\subsection*{Transport measurements}\label{sec:transport}
The arrangement of transport leads for our measurements is shown in Fig. S\ref{fig:transport}a. We measure the momentum relaxing mean free paths with leads 2 and 3, which are separated by approximately 7 $\mu$m. A measurement of the mean free path (mfp) across leads 2 and 3 is shown in Fig.S\ref{fig:transport}b. Figure S\ref{fig:transport}c shows the voltage as a function of source current in our device and displays a linear response. Figure S\ref{fig:transport}d shows the voltage drop across the slit as a function of the gate voltage on the back gate for an AC current of $150\mathrm{\mu A}$. The double-peaked structure results from the source-drain voltage bias shifting the charge density differently for positive and negative currents. We account for this by modulating the gate voltage at the modulation frequency of the current with an amplitude determined by half the peak-peak distance.

\subsection*{Stray
electric fields in a graphene constriction}
\begin{figure*}
\begin{centering}
\includegraphics{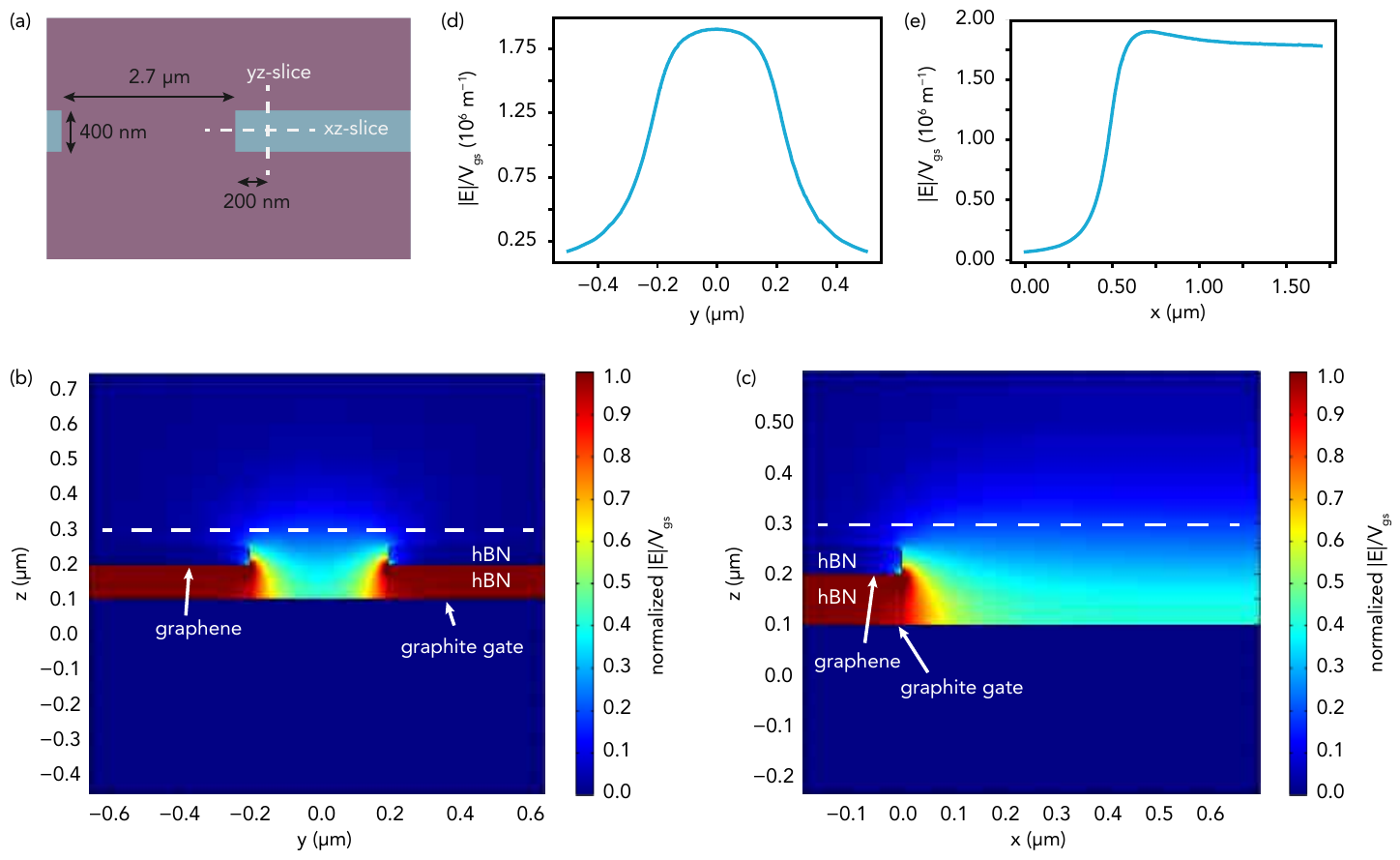}
\let\nobreakspace\relax
\caption{\label{fig:constriction_Efield} Calculation of the stray electric produced by gating a graphene device in a constriction geometry. (a) Birds-eye diagram of the simulated constriction geometry. White dashed line indicate the positions of the slices through the calculated electric fields shown in (b) and (c). (b) $yz$ slice of calculated electric field. (c) $xz$ slice of calculated electric field. (d,e) Line-cuts at the approximate NV scan height (100 nm above graphene) at positions of the white dashed lines in (b) and (c) respectively.}
\end{centering}
\end{figure*}

\begin{figure*}
\begin{centering}
\includegraphics{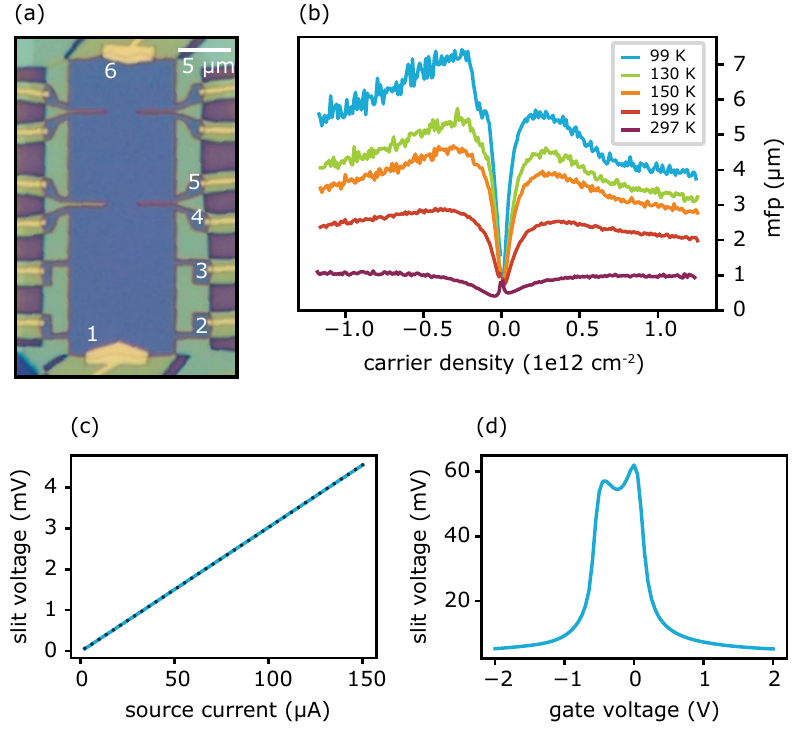}
\let\nobreakspace\relax
\caption{\label{fig:transport} (a) Optical image of device with transport leads labeled. Excitation currents are driven from lead 1 to lead 6. The voltage across leads 2 and 3 is used to measure $\ell_{\mathrm{mr}}$, as Fig. \ref{fig:fig3}g of the main text. (d) Transport measurements \textit{in-situ} of the device. The mean free path as measured by transport supplies an additional means to constrain $\ell_{\mathrm{mr}}$. The excitation current at 79 Hz was 100 nA at 297 K and 500 nA at all other temperatures. (c) Linearity of device tested up to 150 $\mu$A at $n=5$e11/cm$^2$ and $T=123$ K. Source current at 79 Hz flows from lead 1 to lead 6 and  leads 4 and 5 are used for measurement of the constriction voltage. (d) Measurement of the source bias-dependent CNP at $T=123$ K. An excitation current of 150 $\mu$A at 79 Hz results in a double-peaked gate sweep.}
\end{centering}
\end{figure*}

Electric fields at the edges of the etched graphene device can affect the scanning measurements of current density in two ways. First, the stray electric fields resulting from a finite gate-graphene potential difference can be large enough at the position of the NV to shift the spin transition frequencies. In our measurements, the strain is small, the perpendicular magnetic field $B_\perp$ is much smaller than the NV crystal field splitting $D$, $B_\perp \lessapprox B_{\mathrm{NV}}$, the frequency coupling of the perpendicular electric is much smaller than coupling to the NV-axis external magnetic field $d_\perp E_\perp/\gamma B_z \ll 1$, and in this case shifts in transition frequencies due to an external electric field become \cite{Dolde2011},
\begin{equation}
\Delta f_{+/-} \simeq d_{||} E_z \pm  \frac{\left(d_\perp E_\perp\right)^2}{\gamma |B_z|}c
\end{equation}
where $d_\perp=17$ Hz cm/V and $d_{||}=0.35$ Hz cm/V. The differential $\pm I$ current measurements used during scanning lead to the cancellation of most of this frequency shift due to the stray electric fields. However, due to the large source biases used to drive 150 $\mu$A, there is a frequency remnant shift caused by the changing source voltage at the constriction location $V_{\mathrm{s}}$. Defining $\boldsymbol{\alpha}$ as the stray electric field normalized by the gate-graphene potential difference, $\mathbf{E}=\boldsymbol{\alpha} (V_{\mathrm{g}}-V_{\mathrm{s}})$, the remnant frequency shifts due to parallel and perpendicular electric fields are
\begin{align}
\label{eqn:Efield_shift_para}
\Delta f_{||} &\approx 2 d_{||} \alpha_{\mathrm{NV}} V_{\mathrm{s}} \\
\label{eqn:Efield_shift_perp}
\Delta f_{\perp} &\approx 4 \frac{\alpha_\perp^2 d_{\perp}^2}{\gamma |B_{\mathrm{NV}}|} V_{\mathrm{g}} V_{\mathrm{s}}
\end{align}
Figure S\ref{fig:constriction_Efield} shows the results of a finite element electrostatic analysis used to estimate the magnitude of these electric field frequency shifts. For carrier densities $n>5$e11/cm$2$, the typical biases at the constriction $V_{\mathrm{s}}$ result in remnant frequency shifts $<10$ kHz. For comparison, the typical noise in the NV frequency measurement in these scans is greater than $30$ kHz. This shows that the stray electric field is likely not giving large systematic errors in the measured stray magnetic field at the constriction. However, fringing electric field effects may be affecting the measurement in another way--- these fringe fields are expected to affect the distribution of carrier density in the device \cite{Silvestrov2008}. This effect provides another source of gate-dependent changes in the measured magnetic field near the edge of a constriction or channel.

\begin{equation}
    \ell_{\mathrm{mfp}} = \frac{\sigma h}{2e^2 k_{\mathrm{F}}}
\end{equation}

\newpage
\section*{Numerical simulations of Boltzmann transport theory}
We solve the Boltzmann equation within linear response, for particles moving on a circular Fermi surface in two spatial dimensions, using the relaxation time approximation.   While this approximation is imperfect in two dimensional Fermi liquids with parametrically long $\ell_{\mathrm{mr}}$ and with $T/\epsilon_{\mathrm{F}} \ll 1$ \cite{Ledwith2019}, it is a reasonable model for the onset of hydrodynamic effects in materials where electron-electron scattering is within an order of magnitude  of momentum relaxing scattering rates and $T/T_{\mathrm{F}} \sim 0.2$ \cite{KrishnaKumar2017, bandurin_fluidity_2018}.

Let us now review the linearized kinetic theory.   We may approximate the distribution function $f(\mathbf{x},\mathbf{p})$ for electrons at position $\mathbf{x}$ and momentum $\mathbf{p}$ to be \begin{equation}
    f(\mathbf{x},\mathbf{p}) = \Theta(p_{\mathrm{F}}-|\mathbf{p}|) + \delta(p_{\mathrm{F}}-|\mathbf{p}|) \frac{\Phi(\mathbf{x},\theta)}{v_{\mathrm{F}}} + \cdots
\end{equation}
where $\Phi$ denotes the linear perturbation of the Fermi surface due to the applied current (due to the $\delta$ function this term need only be evaluated along the circle  $|\mathbf{p}| = p_{\mathrm{F}}$, in terms of the angular coordinate $\tan\theta = p_y/p_x$), and the $\cdots$ denotes terms which are nonlinear in the applied current; we assume such terms are small.   This expression becomes exact at very low temperatures, but is a reasonable approximation at finite temperature as well.

In terms of the applied electric field $\mathbf{E}$, the time independent linear Boltzmann equation reads, in the absence of a constriction, \begin{equation}
\cos \theta \partial_x \Phi + \sin \theta \partial_y \Phi + \mathsf{W}(\Phi) = -e(E_x  \cos\theta +  E_y \sin\theta)  \label{eq:boltzmann}
\end{equation}
where $\mathsf{W}(\Phi)$ denotes the linearized collision integral within our relaxation time approximation.  To write the explicit form for this collision integral -- and for the manipulations that follow -- it is useful to write \begin{equation}
    \Phi(\mathbf{x},\theta) = \sum_{m=-\infty}^\infty \Phi_m(\mathbf{x}) \mathrm{e}^{\mathrm{i}m\theta}.
\end{equation}
The Boltzmann equation is an infinite dimensional linear equation, and we may employ the Dirac notation (following textbook quantum mechanics)  to describe the angular harmonics $\Phi_m$: we write \begin{equation}
  \Phi(\theta) \rightarrow  |\Phi\rangle = \sum_{m=-\infty}^\infty \Phi_m|m\rangle.
\end{equation}
In the Dirac notation, the linearized collision integral $\mathsf{W}(\Phi)= \mathsf{W}|\Phi\rangle$, where the matrix \begin{equation}
    \mathsf{W} = \frac{1}{\ell_{\mathrm{ee}}} \left(1 - |0\rangle\langle 0| \right) + \left(\frac{1}{\ell_{\mathrm{mr}}} - \frac{1}{\ell_{\mathrm{ee}}}\right)\left(|1\rangle\langle 1| + |-1\rangle\langle -1|\right).
\end{equation}
(Here 1 denotes the identity matrix.)   We also write \begin{equation}
    \mathsf{L} = \cos\theta \partial_x + \sin\theta \partial_y = \sum_{m=-\infty}^\infty\left[ (\partial_x + \mathrm{i}\partial_y)|m\rangle\langle m-1| + (\partial_x - \mathrm{i}\partial_y)|m\rangle\langle m+1| \right].
\end{equation}
Lastly, we will define the vectors \begin{subequations}\begin{align}
    |X\rangle &= \frac{|1\rangle + |-1\rangle}{\sqrt{2}}, \\
    |Y\rangle &= \frac{|1\rangle + |-1\rangle}{\sqrt{2}\mathrm{i}}.
\end{align}\end{subequations}
Observe that the two components of the electrical current density are given by \begin{equation}
    j_x = c \langle X|\Phi\rangle, \;\;\;\;\;\;\; j_y = c \langle Y|\Phi\rangle. 
\end{equation}
The constant of proportionality $c$ is not relevant for our imaging experiments.

We solve these equations in a computational domain \begin{equation}
  -\frac{L_x}{2} < x \le \frac{L_x}{2},\;\;\;\;\;\;\;  -\frac{L_y}{2} < y \le \frac{L_y}{2}
\end{equation}
and impose periodic boundary conditions for computational convenience.
In the absence of a constriction, an exact solution to the Boltzmann equation is \begin{equation}
    |\Phi_0\rangle = -\frac{e E_y \ell_{\mathrm{mr}}}{\sqrt{2}} |Y\rangle.
\end{equation}
Without loss of generality, we will choose $E_y$ so that the coefficient of proportionality above is 1; we will then evaluate the total current flowing in the $y$ direction to properly rescale the solution.

Now we modify these equations to include a constriction using the simple algorithm of \cite{Guo2017}, which imposes a particular set of boundary conditions, roughly similar to ``no slip" boundary conditions which pin the fluid velocity to zero at the edges of the constriction.  (This assumption is likely not exact in graphene; however, we expect the flow through the constriction is much less sensitive to these boundary conditions than other geometries like narrow channels.)  With the fundamental domain above, we define the function \begin{equation}
    \Theta_{\mathrm{c}}(x,y) = \left\lbrace \begin{array}{ll} 1 &\ |x| > \frac{1}{2}w, \; |y|<\frac{t}{2} \\ 0 &\ \text{otherwise} \end{array} \right..
\end{equation}
This function is a step function which equals 1 inside of the constriction, and 0 outside the constriction; the constriction itself is of thickness $t$ in the $y$-direction and has an opening of width $w$ in the $x$-direction.  We then modify (\ref{eq:boltzmann}) to \begin{equation}
    (\mathsf{W}+\mathsf{L})|\Phi\rangle = \frac{\sqrt{2}}{\ell_{\mathrm{mr}}} |Y\rangle - \mathsf{B}|\Phi\rangle 
\end{equation}
where \begin{equation}
  \mathsf{B}=  \alpha \Theta_{\mathrm{c}}(x,y) (|X\rangle\langle X| + |Y\rangle\langle Y|)
\end{equation}
Here $\alpha \sim 10^8 / \ell_{\mathrm{ee}}$ is a very large computational parameter, which forces the flow to halt at the edges of the constriction.

We solve this equation numerically as follows.  First, we write \begin{equation}
    |\Phi\rangle = |\Phi_0\rangle + |\varphi\rangle, \label{eq:phidefs}
\end{equation}
where $|\varphi\rangle \ne 0$ because of the presence of the constriction (implemented by the $\mathsf{B}$ term).  Then \begin{equation}
    (\mathsf{W}+\mathsf{L}+\mathsf{B})|\varphi\rangle = -\mathsf{B}|\Phi_0\rangle.  \label{eq:WLB}
\end{equation}
Second, it is useful to write the abstract vector $|\Phi\rangle$ in two pieces: one outside the constriction and one inside: \begin{equation}
   |\Phi\rangle= \left(\begin{array}{c} |\Phi((x,y)\text{ inside constriction})\rangle \\ |\Phi((x,y)\text{ outside constriction})\rangle \end{array}\right) = \left(\begin{array}{c} |\Phi\rangle_{\mathrm{I}} \\ |\Phi\rangle_{\mathrm{O}} \end{array}\right).
\end{equation} Defining \begin{equation}
    \mathsf{G} = (\mathsf{W}+\mathsf{L})^{-1} = \left(\begin{array}{cc} \mathsf{G}_{\mathrm{II}} &\ \mathsf{G}_{\mathrm{IO}} \\ \mathsf{G}_{\mathrm{OI}} &\ \mathsf{G}_{\mathrm{OO}} \end{array}\right)
\end{equation} (\ref{eq:WLB}) becomes \begin{equation}
    \left(\begin{array}{c} |\varphi\rangle_{\mathrm{I}} \\ |\varphi\rangle_{\mathrm{O}} \end{array}\right) + \left(\begin{array}{c} \mathsf{G}_{\mathrm{II}} \mathsf{B} |\varphi\rangle_{\mathrm{I}} \\ \mathsf{G}_{\mathrm{OI}} \mathsf{B}|\varphi\rangle_{\mathrm{I}} \end{array}\right)= -\left(\begin{array}{c} \mathsf{G}_{\mathrm{II}} \mathsf{B} |\Phi_0\rangle_{\mathrm{I}} \\ \mathsf{G}_{\mathrm{OI}} \mathsf{B}|\Phi_0\rangle_{\mathrm{I}} \end{array}\right)
\end{equation}
which is solved by (upon using (\ref{eq:phidefs})) \begin{subequations} \label{eq:Phisolution}
\begin{align}
    |\Phi\rangle_{\mathrm{I}} &= (1 + \mathsf{G}_{\mathrm{II}}\mathsf{B})^{-1} |\Phi_0\rangle_{\mathrm{I}},  \\
    |\Phi\rangle_{\mathrm{O}} &= |\Phi_0\rangle_{\mathrm{O}} -\mathsf{G}_{\mathrm{OI}}\mathsf{B} |\Phi\rangle_{\mathrm{I}}, 
\end{align}
\end{subequations}

A key feature of these equations is as follows: since $|\Phi_0\rangle = |Y\rangle$ and $\mathsf{B}$ projects onto $|X\rangle$ and $|Y\rangle$, we can, in (\ref{eq:Phisolution}), restrict $|\Phi\rangle$ and $\mathsf{G}$ to the $|X\rangle$ and $|Y\rangle$ sector.  Noting that $\mathsf{G}_{\mathrm{II}}$ and $\mathsf{G}_{\mathrm{OI}}$ simply correspond to restrictions of $\mathsf{G}(x,y; x^\prime, y^\prime)$ (a Green's function in both Fermi surface harmonics and spatial position) to points where $(x,y)$ and $(x^\prime,y^\prime)$ lie inside or outside the constriction, it suffices to calculate $\mathsf{G}(x,y)$ in the absence of a constriction.  In the Fourier domain, one finds \cite{lucas_stokes_2017} \begin{subequations}\begin{align}
    \langle X|\mathsf{G}(k_x,k_y)|X\rangle &= \frac{k_y^2}{k_x^2+k_y^2} \mathcal{G}(k), \\
    \langle X|\mathsf{G}(k_x,k_y)|Y\rangle=\langle Y|\mathsf{G}(k_x,k_y)|X\rangle &= -\frac{k_yk_x}{k_x^2+k_y^2} \mathcal{G}(k), \\
    \langle Y|\mathsf{G}(k_x,k_y)|Y\rangle &= \frac{k_x^2}{k_x^2+k_y^2} \mathcal{G}(k), 
\end{align}
\end{subequations}
where $k=\sqrt{k_x^2+k_y^2}$ and \begin{equation}
    \mathcal{G}(k) = \frac{2\ell_{\mathrm{ee}}\ell_{\mathrm{mr}}}{2\ell_{\mathrm{ee}} - \ell_{\mathrm{mr}} + \ell_{\mathrm{mr}}\sqrt{1+k^2\ell_{\mathrm{ee}}^2}}
\end{equation}
We numerically invert this Fourier transform to calculate $\mathsf{G}_{\mathrm{II}}$ and $\mathsf{G}_{\mathrm{OI}}$, and dealias to reduce spectral noise in the real space result.  It is then straightforward to solve (\ref{eq:Phisolution}).  We emphasize that when the constriction is a relatively small part of the computational domain, the only matrix inverse in (\ref{eq:Phisolution}) remains relatively small, allowing us to use $> 150$ of grid points in the $x$ direction and $> 50$ grid points in the $y$ direction and solve these kinetic equations easily on a personal computer.

\newpage
\section*{Comparison between the channel and slit geometries}

\begin{figure*}
\begin{centering}
\includegraphics[width = 6 in]{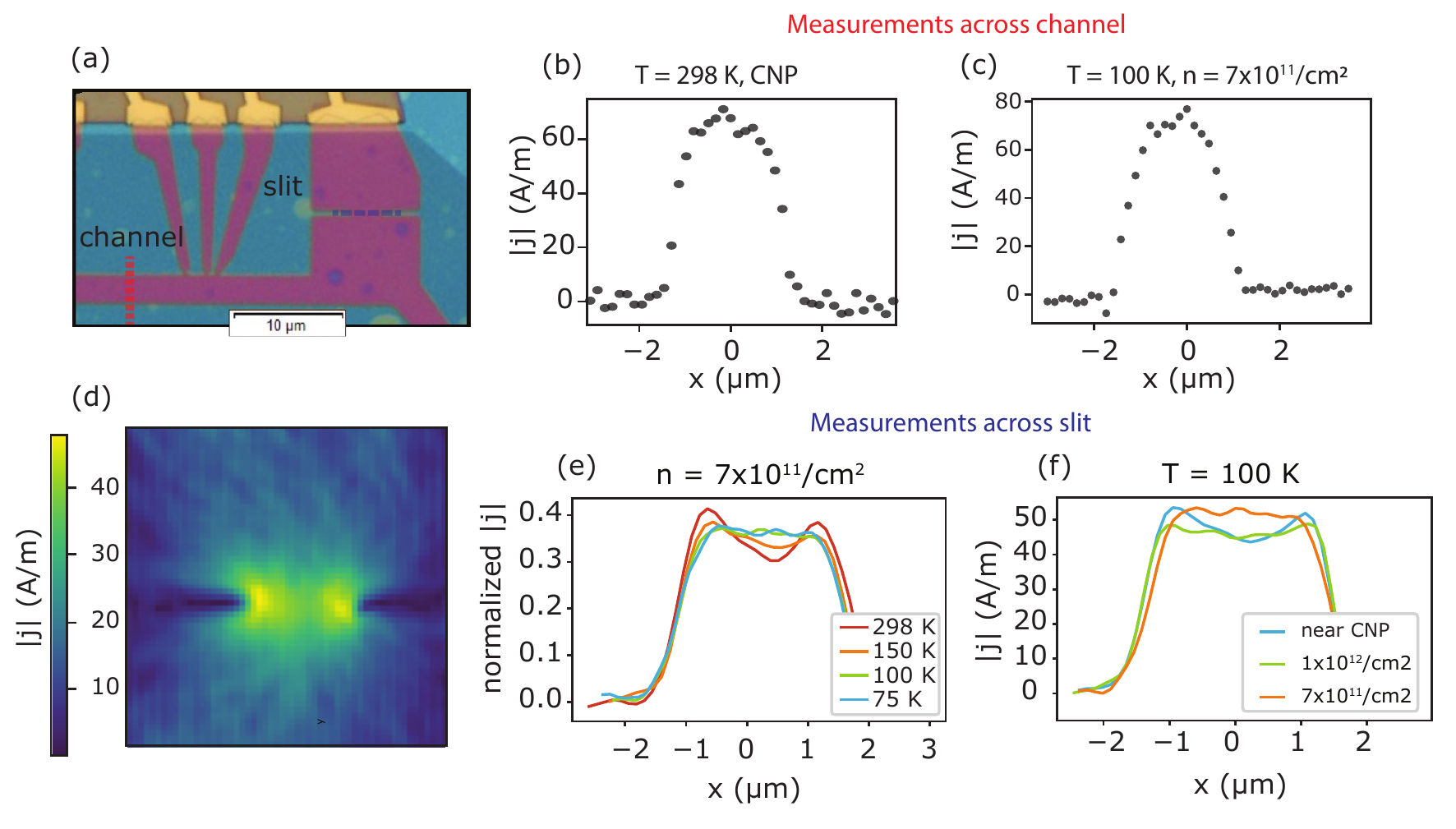}
\let\nobreakspace\relax
\caption{\label{fig:channel} Measurements of the current density flowing along a 2.7 $\mu$m wide channel. (a) Optical image of the secondary graphene device showing the locations used in obtaining the current density measurements for the channel geometry (b-c) and the slit geometry (d-f). (b) Current density profile in the channel near the CNP at 298 K. The black dots are the reconstructed current density. (c) Measurement of the current density profile of the channel at the same position as in (b), at $T=100$ K and $n=7$e11/cm$^2$. The NV used to measure (b) and (c) was oriented at $\theta=58^\circ$ and $\phi=28^\circ$ relative to the scanning line. (d) Reconstructed current density magnitude $|j|$ at 298 K, near the CNP, showing the characteristic double peaks of ohmic flow. (d) Temperature dependence of the reconstructed $j_y$ component of current density profile at fixed carrier density $n=\SI{7e11}{/\centi\meter^2}$ in a linecut through the constriction. (e) Carrier density dependence of the $j_y$ profile at fixed temperature $T=100$ K. The NV used to measure (d) was oriented at $\theta=58^\circ$ and $\phi=108^\circ$ relative to the scanning line in (a), while the NV used in the measurements (e--f) was oriented at $\theta=53^\circ$ and $\phi=-71^\circ$ relative to the same axes.}
\end{centering}
\end{figure*}

In addition to the slit geometry, we also investigate the channel geometry using a similar device to the one described in the main text. In contrast to the slit geometry, in the channel geometry we assume that the current is slowly varying in directions orthogonal to the scan-axis in order to reconstruct the current. We pattern both a slit and a channel on this device to investigate the profiles of both at room temperature and low temperatures at both CNP and finite charge densities. We compare our results to those obtained in Ref. \cite{Ku2019}. Figure S\ref{fig:channel}a shows an optical image of the device used for comparative measurements between the channel and the slit geometry. This device consists of a ML of graphene encapsulated between upper and lower hBN layers with thicknesses of 52 nm and 61 nm respectively. We use a graphite gate to tune the charge carrier density. At both high and low temperatures we are able to obtain profiles that are approximately parabolic and are consistent with the expected parabolic profiles for viscous Poiseuille flow. Figure \ref{fig:channel}b-c shows the result of line profiles at room-temperature near the CNP and at low temperature at finite charge density. We find that the shape of the profiles are not significantly different between the room-temperature data at CNP and the low-temperature data at finite charge density. Because the competing scattering lengths are expected to change both as a function of temperature and charge density, we conclude that the channel geometry is only subtly sensitive to changes in the scattering lengths. Similar profiles to these are used in Ref. \cite{Ku2019} to claim the existence of a hydrodynamic regime at room-temperature and close to the CNP. 

The slit in the secondary device has dimensions $w=\SI{2.7}{\micro\meter}\times t=\sim \SI{400}{\nano\meter}$. Figure S\ref{fig:channel}d shows the reconstructed current density magnitude $|j|$ at $T=298$ K and near the CNP. We observe the double-peaked structure characteristic of ohmic transport, indicating that the device is locally ohmic at these temperatures and charge density. Figure \ref{fig:channel} e-f show linecuts of the current across the slit for the device described in this section and reveal decidedly ohmic behavior at all charge densities studied at room temperature measurements. At lower temperatures and finite charge density, we observe the disappearance of the ohmic double peaks for the secondary device, reproducing the results of the primary device described in the main text of the paper. Our measurements on the secondary device support the conclusions made in main text, contrasting sharply with the measurements on the channel and those made by Ref. \cite{Ku2019}.

The authors reconcile the ostensible contradiction between the measurements on the slit and the channel geometries by noting that the slit is much less sensitive to the relatively unknown state of the etched device boundaries due to the slit edge being small compared to the mean free path. Due to the lack of sensitivity to the boundary conditions and dramatic ohmic signature, the slit is better suited to resolve the transition between locally ohmic transport and viscous flows. We attribute the rounded shape of the current profiles in the channel geometry to poorly-understood boundary-enhanced scattering processes, for example: bound charges shifting the local effective potential close to the edges. Without a comprehensive understanding of scattering at the device edges, the results on the channel are difficult to draw conclusions from, a fact that re-asserts the suitability of the slit geometry for measuring the ohmic-viscous crossover.

\end{document}